\documentclass[11pt,a4paper]{article}

\usepackage[title]{appendix}
\usepackage[utf8]{inputenc}
\usepackage[english]{babel}
\usepackage[T1]{fontenc}
\usepackage{amsmath}
\usepackage{amsthm}
\numberwithin{equation}{section}
\usepackage{amssymb}
\usepackage{amsfonts}
\usepackage{amssymb}
\usepackage{gensymb}
\usepackage{graphicx}
\usepackage[left=2.5cm,right=2.5cm,top=3cm,bottom=2.5cm]{geometry}
\usepackage{caption}
\usepackage{subcaption} %package for splittet graphs
\usepackage{siunitx} %package for units
\usepackage{fancyvrb}%package-upgrade for verbatim
\usepackage{multirow} %package for multirows in tables
\usepackage[labelfont={bf}]{caption} %allows global Margins for Captions and others
\usepackage{floatrow} % package allows sideways captions
\usepackage{upgreek}
\usepackage{esvect}
\usepackage{tabularx}
\usepackage[version=3]{mhchem}
\usepackage{bbm}
\usepackage{mathtools}

\DeclareMathOperator{\One}{\mathbbm{1}}

\DeclareMathOperator{\ii}{\textrm{i}}
\usepackage{tikz-feynman} %package for Feynman diagrams
\usepackage{bbold}
\usepackage{listings} %for code display
\usepackage{xcolor} %setting colors
\usepackage[square,numbers]{natbib}

\usepackage{textcomp}   % \textperthousand  (promille)
\usepackage[autostyle=true,german=quotes]{csquotes}
\usepackage{fancyhdr}%settings for header and footer
\usepackage{url}
\usepackage{hyperref} %package for links

\tikzfeynmanset{compat=1.0.0}
\newcommand{\myparagraph}[1]{\paragraph*{#1}\mbox{}\\}
\newcommand\scalemath[2]{\scalebox{#1}{\mbox{\ensuremath{\displaystyle #2}}}}

\newcolumntype{L}[1]{>{\raggedright\arraybackslash}p{#1}} % linksbündig mit Breitenangabe
\newcolumntype{C}[1]{>{\centering\arraybackslash}p{#1}} % zentriert mit Breitenangabe
\newcolumntype{R}[1]{>{\raggedleft\arraybackslash}p{#1}} % rechtsbündig mit Breitenangabe
\title{Locality of staggered overlap operators}
\author{Nuha Chreim\thanks{nuha.chreim@uni-wuppertal.de}, Christian Hoelbling\thanks{hch@uni-wuppertal.de}, and Christian Zielinski\thanks{email@czielinski.de}}

\date{\it{Bergische Universität Wuppertal, Gau{\ss}stra{\ss}e 20, D-42119 Wuppertal, Germany}}

\begin{document}

\maketitle

\begin{abstract}
We give an explicit proof for the locality of staggered overlap operators. The proof covers the original two flavor construction by Adams as well as a single flavor version. As in the case of Neuberger's operator, an admissibility condition for the gauge fields is required.
\end{abstract}

\section{Introduction and motivation}
As Adams has shown ~\cite{Adams:2010gx}, it is possible to construct
chirally symmetric lattice fermions based on the staggered
discretization. While Adams' original construction provided a two
flavor operator, single flavor versions were found soon after
\cite{Hoelbling:2010jw,Hoelbling:2016qfv}. These staggered chiral
fermions are obtained by first adding a mass term
\cite{GOLTERMAN198461} to the staggered operator, followed by an
overlap construction \cite{Neuberger:1998}, which contains an inverse
square root. It is thus evident that staggered chiral fermions are not
ultralocal by construction and their locality needs to be
proven. Numerically, Ref.~\cite{deForcrand:2010wrz} found strong
evidence in support of the locality of Adams' original two flavor
operator.
In the free case, one can furthermore show that the lifting of the doubler modes
is achieved via flavor dependent mass term
\cite{Adams:2010gx,Hoelbling:2010jw,deForcrand:2012bm}.
In addition, the index theorem has been established for the two flavor operator ~\cite{Adams:2009eb} and the correct continuum limit of the index was found in ~\cite{Adams:2013lpa}. In this paper, we give an analytic proof for the locality of staggered overlap fermions, for both the single and two flavor cases. The general strategy we employ  is quite similar to the one used by  Hernández, Jansen and Lüscher to demonstrate the locality of the original Neuberger operator \cite{Hernandez:1998et}. We will start in sec.\,2 by expanding the inverse square root as a series of Legendre polynomials, which can be shown to be local if a spectral condition of the kernel operator is fulfilled. This spectral condition involves an upper as well as a lower bound on the kernel operator. In sec.\,3 we will show that both bounds are fulfilled for Adams' original two flavor construction, provided an admissibility condition of the form $\left\Vert \One-P\right\Vert <\varepsilon$ is fulfilled by all plaquettes $P$ of the gauge field. The exact value of $\varepsilon$ will depend on the details of the action, specifically the negative mass parameter $s$ and the Wilson parameter $r$. We then turn to a single flavor staggered operator and show that similar bounds also hold in this case.

\section{Locality}

\subsection{Staggered overlap Dirac operator}

 Let us first introduce the staggered overlap Dirac operator
\begin{equation}
D_{\mathsf{so}}=\frac{1}{a}\left(\One+A/\sqrt{A^{\dagger}A}\right)
\end{equation}
with 
\begin{equation}
A =aD_{\mathsf{sw}}-rs\One \qquad
D_{\mathsf{sw}} =D_{\mathsf{st}}+W_{\mathsf{st}}
\end{equation}
where $r$ is the Wilson parameter and $0<s<2$ is the negative mass
term of the kernel operator. The staggered operator is defined as
\begin{equation}
D_{\textrm{st}}=\eta_{\mu}\nabla_{\mu}
\end{equation}
with
\begin{equation}
(\eta_{\mu})_x=(-1)^{\sum_{\nu < \mu}x_{\nu}}
\end{equation}
and the symmetric derivative operator
\begin{equation}
\nabla_{\mu}=\frac{1}{2a}\left(T_{\mu+}-T_{\mu-}\right). \label{Nabla-mu}
\end{equation}
The $T_{\mu\pm}$ are parallel transports defined as
\begin{equation}
  (T_{\mu+})_{xy}=U_{\mu}(x)\delta_{x+\hat{\mu},y}
                   \qquad
(T_{\mu-})_{xy}=U_{\mu}^{\dagger}(y)\delta_{x-\hat{\mu},y}.
\end{equation}
The staggered Wilson term $W_{\mathsf{st}}$  reads 
\begin{equation}
W_{\mathsf{st}}=\frac{r}{a}\left(\One-M^{(\mathsf {2})}\right)
\end{equation}
in the two flavor case \cite{Adams:2010gx,Adams:2009eb,Adams:2011xf}
and 
\begin{equation}
W_{\mathsf{st}}=\frac{r}{a}\left(2\cdot\One+M^{(\mathsf{1})}\right)
\end{equation}
in the one flavor case \cite{Hoelbling:2010jw,Hoelbling:2016qfv}. The operators
$M^{(\mathsf{N_f})}$ are in turn given\footnote{
  Note that in principle more general single flavor terms are allowed
  \cite{Hoelbling:2016qfv}. These are, however, not substantially different
  and the generalization is straightforward.
}
by
\begin{equation}
  M^{(\mathsf {2})}=\epsilon \eta_5 C
  \qquad
  M^{(\mathsf {1})}=\ii\eta_{12}C_{12}+\ii\eta_{34}C_{34} 
\end{equation}
with the phase factors
\begin{align}
\eta_5&=\eta_1\eta_2\eta_3\eta_4 \\
\epsilon_x&=(-1)^{\sum_{\nu}x_{\nu}}\\
(\eta_{\mu\nu})_x &=-\eta_{\nu\mu}=(-1)^{\sum_{\rho=\mu+1}^{\nu}x_{\rho}}\textrm{ for } \mu\le\nu                   
\end{align}
and the diagonal hopping terms
 \begin{align}
 C=(C_1C_2C_3C_4)_{\text{sym}}&=\frac{1}{4!}P_{\alpha \beta \gamma \delta}C_{\alpha}C_{\beta}C_{\gamma}C_{\delta}\\
  C_{\mu\nu} & =\frac{1}{2}\left\{ C_{\mu},C_{\nu}\right\} 
\end{align}
where
\begin{equation}
  C_{\mu}=\frac{1}{2}(T_{\mu+}+T_{\mu-})\label{Cmu}
\end{equation}
and $P_{\alpha \beta \gamma \delta}$ denotes the
permutation symbol
\begin{equation}
  P_{\alpha \beta \gamma \delta}=\left\{
      \begin{array}{lcl}
        1&&\alpha,\beta,\gamma,\delta \text{ is a permutation of } 1,2,3,4\\
        0&& \text{else.}
        \end{array}
    \right.
\end{equation}

The kernel $A$ is ultralocal, but due to the $\left(A^{\dagger}A\right)^{-1/2}$
term the staggered overlap Dirac operator $D_{\mathsf{so}}$ is not.
However, if the matrix elements $(D_{\mathsf{so}})_{x,y}$ of the
staggered overlap operator 
are decaying exponentially for large distances $\left\Vert x-y\right\Vert $
with a decay constant $\propto a^{-1}$, then we recover a local
operator in the continuum limit.

\subsection{Legendre series expansion}

Following the strategy employed in Ref.\,\cite{Hernandez:1998et}
we begin by expanding $\left(A^{\dagger}A\right)^{-1/2}$ in a series
of Legendre polynomials. In order to make the expansion convergent
we impose the following inequality, which we will show in sect.\,\ref{sec:Bounds}:
\begin{equation}
0<u\leq A^{\dagger}A\leq v<\infty.\label{eq:MasterInequality}
\end{equation}
The inequality stands for the corresponding inequality between the expectation values of
the operators in arbitrary normalizable states. We also explicitly assume that $u<v$. In the following we can set $u=\lambda_{\min}$ and $v=\lambda_{\max}$ as noted in Ref.\,\cite{Adams:2003rr}.

The Legendre polynomials $P_{k}\left(z\right)$ can be defined through
the expansion of the generating function
\begin{equation}
\left(1-2tz+t^{2}\right)^{-1/2}=\sum_{k=0}^{\infty}t^{k}P_{k}\left(z\right).\label{eq:LegendreExpansion}
\end{equation}
We can now set
\begin{equation}
z=\frac{\left(\lambda_{\min}+\lambda_{\max}\right)\One-2A^{\dagger}A}{\lambda_{\max}-\lambda_{\min}}
\end{equation}
and due to eq.\,\eqref{eq:MasterInequality} find that this operator
has norm $\left\Vert z\right\Vert=1$. Here and in the following $\left\Vert \cdot\right\Vert =\left\Vert \cdot\right\Vert _{2}\equiv\sigma_{\max}\left(\cdot\right)$
refers to the spectral norm and $\sigma_{\max}$ refers to the largest
singular value.

Then the property $\left|P_{k}\left(x\right)\right|\leq1$  $\forall x\in\left[-1,1\right]$
together with $\left\Vert z\right\Vert = 1$ translates to
\begin{equation}
\left\Vert P_{k}\left(z\right)\right\Vert \leq1. \label{eq:LegendreNorm}
\end{equation}
It follows that eq.\,\eqref{eq:LegendreExpansion} is norm convergent
for our choice of $z$ for all $t$ satisfying $\left|t\right|<1$.
Due to eq.\,\eqref{eq:MasterInequality}, we can now introduce $\theta$ through
\begin{equation}
\cosh\theta=\frac{\lambda_{\max}+\lambda_{\min}}{\lambda_{\max}-\lambda_{\min}},\qquad\theta>0,
\end{equation}
and set
\begin{equation}
t=e^{-\theta},
\end{equation}
which implies $0<t\le 1$, so that the series is convergent. Note that
this allows us to express $t$ as
\begin{equation}
t=\cosh\theta-\sqrt{\cosh^2\theta-1}=\frac{\sqrt{\lambda_{\max}}-\sqrt{\lambda_{\min}}}{\sqrt{\lambda_{\max}}+\sqrt{\lambda_{\min}}}.
\end{equation}
From eq.\,\eqref{eq:LegendreExpansion} we thus obtain
\begin{align}
\left(1-2tz+t^{2}\right)^{-1/2} & =\left(1-\frac{2t}{\lambda_{\max}-\lambda_{\min}}\left(\lambda_{\min}+\lambda_{\max}-2A^{\dagger}A\right)+t^{2}\right)^{-1/2}\nonumber \\
 & =\left(1-2e^{-\theta}\cosh\theta+\frac{4t}{\lambda_{\max}-\lambda_{\min}}A^{\dagger}A+e^{-2\theta}\right)^{-1/2}\nonumber \\
 & =\sqrt{\frac{\lambda_{\max}-\lambda_{\min}}{4t}}\left(A^{\dagger}A\right)^{-1/2}\nonumber \\
 & =\frac{\sqrt{\lambda_{\max}}+\sqrt{\lambda_{\min}}}{2}\left(A^{\dagger}A\right)^{-1/2}
\end{align}
and therefore
\begin{equation}
\left(A^{\dagger}A\right)^{-1/2}=\kappa\sum_{k=0}^{\infty}t^{k}P_{k}\left(z\right)\label{eq:ExpansionAdagA}
\end{equation}
with
$\kappa=2/\left(\sqrt{\lambda_{\max}}+\sqrt{\lambda_{\min}}\right)$.

\subsection{Locality of the inverse square root}

The lack of ultralocality stems from the $\left(A^{\dagger}A\right)^{-1/2}$
term, hence it is sufficient to establish the locality of that term
in the sense defined earlier. We start by defining the kernel $G\left(x,y\right)$ via
\begin{equation}
  G\left(x,y\right)=
  \left((A^{\dagger}A)^{-1/2}\right)_{xy}.
\end{equation}
Similarly, we define the kernels of the $P_{k}\left(z\right)$
via
\begin{equation}
G_{k}\left(x,y\right)
  =
  \left(P_{k}(z)\right)_{xy}\label{eq:LegendreKernel}
\end{equation}
and use eq.\,(\ref{eq:ExpansionAdagA}) to obtain
\begin{equation}
G\left(x,y\right)=\kappa\sum_{k=0}^{\infty}t^{k}G_{k}\left(x,y\right).
\end{equation}
The norm convergence of the Legendre expansion implies the absolute
convergence of this series for all $ x$ and $y$. From
eq.\,\eqref{eq:LegendreNorm} and eq.\,\eqref{eq:LegendreKernel} we infer that
\begin{equation}
\left\Vert G_{k}\left(x,y\right)\right\Vert \leq1,\qquad\forall x\,\forall y\,\forall k,
\end{equation}
where the norm is in color space.

Because $P_{k}\left(z\right)$ is a polynomial in $A^{\dagger}A$
and $A$ is an ultralocal operator, we find that $G_{k}\left(x,y\right)$
vanishes unless $x$ and $y$ are sufficiently close to each other.
If we introduce the Manhattan distance $\left\Vert \cdot\right\Vert _{1}$,
we have
\begin{equation}
G_{k}\left(x,y\right)=0,\qquad\forall k<\frac{1}{2\ell a}\left\Vert x-y\right\Vert _{1},
\end{equation}
where $\ell$ is the range of the operator $A$ in lattice
units, i.e., the maximum Manhattan distance in lattice units between
points coupled by the operator. For two flavor staggered Wilson fermions we have $\ell=4$, for one
flavor staggered Wilson fermions $\ell=2$ and for Wilson fermions
$\ell=1$. Using the shorthand notation $d= \left\Vert x-y\right\Vert _{1}/(2\ell a) $
we find
\begin{align}
\left\Vert G\left(x,y\right)\right\Vert  & =\kappa\sum_{k=d}^{\infty}t^{k}\left\Vert G_{k}\left(x,y\right)\right\Vert \nonumber \\
 & \leq\kappa\sum_{k=d}^{\infty}t^{k}\nonumber \\
 & =\frac{\kappa}{1-t}t^{d}\nonumber \\
 & =\frac{\kappa}{1-t}\exp\left(-\frac{\theta}{2\ell a}\left\Vert x-y\right\Vert _{1}\right)\nonumber \\
 & = \frac{1}{\sqrt{\lambda_{\min}}}\exp\left(-\frac{1}{\xi}\left\Vert x-y\right\Vert _{1}\right)
\end{align}
and thus an exponential falloff with the decay constant
\footnote{Note that the $\log$ term may provide subleading corrections
to this behavior only.}
\begin{equation}
\xi^{-1}  =\frac{\theta}{2\ell a}=\frac{1}{2\ell a}\log\left(\frac{\sqrt{\lambda_{\max}}+\sqrt{\lambda_{\min}}}{\sqrt{\lambda_{\max}}-\sqrt{\lambda_{\min}}}\right)\propto\frac{1}{a}.
\end{equation}
This establishes the locality of $\left(A^{\dagger}A\right)^{-1/2}$ providing
eq.~\eqref{eq:MasterInequality} holds with the spectral bounds given
by  $u=\lambda_{\min}$ and $v=\lambda_{\max}$. The equivalent of this particular
form for usual overlap fermions was derived in Ref.\,\cite{Adams:2003rr}.

Let us finally remark that this result can be slightly generalised in
the case of a single isolated zero or near zero mode
$\lambda_{\min}$. As shown in sect.~2.4 of Ref.\,\cite{Hernandez:1998et}, one can treat a single isolated zero or near zero mode
separately and still establish locality. In that case we identify the
lower spectral bound $u=\lambda_2$  with the second smallest eigenvalue of $A^\dag A$. If $\lambda_{\min}<u/2$, locality can again be
established \cite{Hernandez:1998et}.

\section{Bounds on $A^{\dagger}A$ \label{sec:Bounds}}

We now need to establish the spectral bounds as defined in eq.\,\eqref{eq:MasterInequality}
for the kernel operator. We first derive some useful identities
and then establish the upper bound, which is straightforward. The
main task is then to establish the lower bound, which we do separately
for the two and one flavor case. In both instances, the bound can be
established given an admissibility condition for the gauge fields.

\subsection{Some useful identities}

We first note that the parallel transports fulfill the relations
$T_{\mu-}=T_{\mu+}^{\dagger}=T_{\mu+}^{-1}$, which implies that the
$T_{\mu \pm}$ are unitary and thus have singular values 1,
i.e. $\Arrowvert T_{\mu \pm}\Arrowvert=1$. The covariant second
derivative operator is given by
\begin{equation}
\Delta_{\mu}=T_{\mu+}+T_{\mu-}-2,
\end{equation}
so we can recast eq.\,\eqref{Cmu} as
\begin{equation}
C_{\mu}=1+\frac{\Delta_{\mu}}{2}.
\end{equation}
Using this relation we find 
\begin{equation}
C_{\mu}^2=1+\frac{1}{4}(T_{\mu+}^2+T_{\mu-}^2-2). \label{definitionC}
\end{equation}
Defining 
\begin{equation}
V_{\mu}=\frac{1}{4}(T_{\mu+}^2+T_{\mu-}^2-2)
\end{equation}
it follows that
\begin{equation}
C_{\mu}^2=1+V_{\mu}.\label{Cmusquared}
\end{equation}
From  eq.\,\eqref{Cmu} we also find
\begin{equation}
\|C_{\mu} \| \leq \frac{1}{2}(\|T_{\mu+}\| +\|T_{\mu-}\|)=1,\label{clt1}
\end{equation}
which implies 
\begin{equation}
\Arrowvert M^{(\mathsf {2})}\Arrowvert=\Arrowvert \eta_5
C\epsilon\Arrowvert \leq 1 \label{eq:M2bound}
\end{equation}
and, since both $\eta_5$ and $\epsilon$ commute with $C$ and square
to the identity, ${M^{(\mathsf {2})}}^2=C^2$. Another useful identity is 
\begin{equation}
a^2\nabla_{\mu}^2= V_{\mu} \label{firstterm}
\end{equation}
which, together with the anti Hermiticity condition
$
\nabla_{\mu}^{\dagger}=-\nabla_{\mu},
$
implies that
\begin{equation}
0\leq a^2\nabla_{\mu}^{\dagger}\nabla_{\mu} =-V_{\mu}. \label{inequalityVmu}
\end{equation}
Additionally, the Hermiticity condition $C_{\mu}^{\dagger}=C_{\mu}$
implies that $C_{\mu}^2 \geq 0$ and thus $1+V_{\mu} \geq 0$.

Next, we want to find a more explicit expressions for
$A^{\dagger}A$. Noting that
\begin{equation}
\nabla_{\mu}\eta_{\nu}=\begin{cases}
\eta_{\nu}\nabla_{\mu} & \mu\geq\nu,\\
-\eta_{\nu}\nabla_{\mu} & \mu<\nu,
\end{cases}
\end{equation}
we find
\begin{equation}
\sum_{\mu,\nu}\eta_{\mu}\nabla_{\mu}\eta_{\nu}\nabla_{\nu} =\nabla^{2}+\sum_{\mu>\nu}\eta_{\mu}\eta_{\nu}\left[\nabla_{\mu},\nabla_{\nu}\right],
\end{equation}
where we have introduced the shorthand notation
\begin{equation}
\nabla^2=\sum_{\mu}\nabla_{\mu}\nabla_{\mu}.
\end{equation}
We then find 
\begin{equation}
  A_{2}^{\dagger}A_{2}=-a^{2}\nabla^{2}-a^{2}\sum_{\mu>\nu}\eta_{\mu}\eta_{\nu}\left[\nabla_{\mu},\nabla_{\nu}\right]+r^{2}\left(\One(1-s)
    -M^{(\mathsf {2})}\right)^{2}-ar\left[M^{(\mathsf {2})},\eta_{\mu}\nabla_{\mu}\right] \label{twoflavor}
\end{equation}
in the two flavor case and
\begin{equation}
A_{1}^{\dagger}A_{1}=-a^{2}\nabla^{2}-a^{2}\sum_{\mu>\nu}\eta_{\mu}\eta_{\nu}\left[\nabla_{\mu},\nabla_{\nu}\right]+r^{2}\left(\One(2-s)+M^{(\mathsf{1})}\right)^{2}+ar\left[M^{(\mathsf{1})},\eta_{\mu}\nabla_{\mu}\right] \label{oneflavor}
\end{equation}
in the one flavor case.

\subsection{Upper bound}
Using $\left\Vert T_{\mu\pm}\right\Vert =1$ we find the following bounds 
\begin{align}
\left\Vert a\nabla_{\mu}\right\Vert  & \leq\frac{1}{2}\left(\left\Vert T_{\mu+}\right\Vert +\left\Vert T_{\mu-}\right\Vert \right)\leq1,\\
\left\Vert a\eta_{\mu}\nabla_{\mu}\right\Vert  & \leq 4, \\
\left\Vert C_{\mu}\right\Vert  & =\frac{1}{2}\left\Vert T_{\mu+}+T_{\mu-}\right\Vert \leq1,\label{cleq1}\\
\left\Vert C\right\Vert  & =\left\Vert \left(C_{1}C_{2}C_{3}C_{4}\right)_{\mathsf{sym}}\right\Vert \leq\frac{1}{4!}\cdot4!\cdot\prod_{\mu}\left\Vert C_{\mu}\right\Vert \leq1,
\end{align}
and using eq.\,\eqref{eq:M2bound} we find 
\begin{equation}
  \left\Vert r\left(
    \One(1-s)-M^{(\mathsf {2})}
  \right)
\right\Vert \leq 
\lvert r \rvert(2-s).
\end{equation}
Putting all this together, we find
\begin{equation}
  \Arrowvert A_2\Arrowvert = \left\Arrowvert
  a\eta_{\mu}\nabla_{\mu}+r\left(
    \One(1-s)-M^{(\mathsf {2})}
\right)\right\Arrowvert\leq 4+\lvert r \rvert(2-s).
\end{equation}
The same bound holds for $A_2^{\dagger}$ and so
\begin{equation}
\Arrowvert A_{2}^{\dagger} A_2\Arrowvert \leq \Arrowvert A_{2}^{\dagger} \Arrowvert \Arrowvert A_2 \Arrowvert \leq (4+\lvert r \rvert(2-s))^2
\end{equation}
is uniformly bounded from above for all $r$ and $s$ and we can establish
the existence of $v$ in eq.~\eqref{eq:MasterInequality} in the
two flavor case.

For the one flavor case we note that
\begin{align}
\left\Vert C_{\mu\nu}\right\Vert  & =\frac{1}{2}\left\Vert \left\{ C_{\mu},C_{\nu}\right\} \right\Vert \leq1,
%\left\Vert M_{\mu\nu}\right\Vert  & =\left\Vert \ii\eta_{\mu\nu}C_{\mu\nu}\right\Vert \leq\left\Vert C_{\mu\nu}\right\Vert \leq1,
\end{align}
from which it follows that
\begin{equation}
\Arrowvert  M^{(\mathsf {1})}\Arrowvert\leq \Arrowvert C_{12}\Arrowvert + \Arrowvert C_{34}\Arrowvert \leq 2.
\end{equation}
Hence we find
\begin{equation}
\left \Arrowvert \One(2-s)+M^{(\mathsf {1})} \right\Arrowvert \leq 4-s
\end{equation}
and it follows, similarly to the two flavor case, that
\begin{equation}
\Arrowvert A_1 \Arrowvert \leq 4+\lvert r \rvert(4-s).
\end{equation}
Since $A_1^\dag$ does obey the same bound, we obtain
\begin{equation}
\Arrowvert A_1^{\dagger}A_1 \Arrowvert \leq (4+\lvert r \rvert(4-s))^2.
\end{equation}
This establishes the existence of $v$ in
eq.\,\eqref{eq:MasterInequality} in the
single flavor case as well.

\subsection{Lower bound}

As $A^{\dagger}A$ is Hermitian and positive semidefinite we are
left with showing the absence of zero-modes. However, in general this
operator can have zero-modes for certain gauge configurations, therefore
no uniform positive lower bound exists. Zero-modes can only be excluded
if we assume the gauge field to be sufficiently smooth. In our case
let us assume that
\begin{equation}
\left\Vert \One-P\right\Vert <\varepsilon\qquad\textrm{for all plaquettes }P.
\end{equation}
As a consequence of the smoothness condition, we obtain the following
relations (see app.~\ref{AppendixA})
\begin{equation}
\left\Vert a^{2}\left[\nabla_{\mu},\nabla_{\nu}\right]\right\Vert <\varepsilon, \label{inequalitynabla}
\qquad
\Arrowvert [C_{\mu},C_{\nu}]\Arrowvert < \varepsilon,
\qquad
\Arrowvert a[C_{\mu},\nabla_{\nu}]\Arrowvert < \varepsilon .
\end{equation}

\subsubsection{Lower bound on the two flavor operator $A_2^{\dagger}A_2$}
There are four terms in 
\begin{equation}
A_{2}^{\dagger}A_{2}=-a^{2}\nabla^{2}-\sum_{\mu>\nu}\eta_{\mu}\eta_{\nu}a^{2}\left[\nabla_{\mu},\nabla_{\nu}\right]+r^{2}\left(\One (1-s)-M^{(2)}\right)^{2}-ar\left[M^{(2)},\eta_{\mu}\nabla_{\mu}\right],
\end{equation}
for which we will find bounds individually. We will consider the case
$0<r\leq 1$ first and derive a bound for $r>1$ later\footnote{The $r<0$ case can be covered by the simple replacement of $r \rightarrow \lvert r \rvert$ in the bounds. However, negative $r$ do not represent a physically different system compared to positive $r$ and will therefore not be considered further.}.

\myparagraph{The first and third term}
We first look at $-a^2\nabla^2+r^2C^2$, where ${M^{(2)}}^2=C^2$ is used. Using inequality \eqref{inequalitynabla} we find  (cf. app.~\ref{AppendixA})
\begin{equation}
\Arrowvert C^2-(C_1^2 C_2^2 C_3^2 C_4^2) _{\text{sym}} \Arrowvert < 9 \varepsilon.
\end{equation}
Using eqs.\,\eqref{Cmusquared}, \eqref{firstterm} and \eqref{inequalityVmu}, we furthermore see that for $0<r\leq 1$
\begin{align}
-a^2\nabla^2+r^2C^2 >& -a^2\nabla^2+r^2 (C_1^2 C_2^2 C_3^2 C_4^2) _{\text{sym}} -9r^2\varepsilon \notag \\
=&-\sum_{\mu}V_{\mu}+r^2\frac{1}{4!} P_{\alpha\beta\gamma\delta}(1+V_{\alpha}) (1+V_{\beta}) (1+V_{\gamma}) (1+V_{\delta})-9r^2\varepsilon \notag \\
  =&-\sum_{\mu}V_{\mu}+r^2+r^2\sum_{\mu}V_{\mu}+\frac{1}{2}r^2\sum_{\mu\neq\nu}V_{\mu}V_{\nu}\notag\\
     &+\frac{1}{3!}r^2\sum_{\mu\neq\nu\neq\alpha\neq\mu}V_{\mu}V_{\nu}V_{\alpha}+r^2(V_1V_2V_3V_4)_{\text{sym}}-9r^2\varepsilon \notag \\
=&r^2-(1-r^2)\sum_{\mu}V_{\mu}+\frac{1}{2}r^2\sum_{\mu\neq\nu}V_{\mu}V_{\nu}\notag\\
     &+\frac{1}{3!}r^2\sum_{\mu\neq\nu\neq\alpha\neq\mu}V_{\mu}V_{\nu}V_{\alpha}+r^2(V_1V_2V_3V_4)_{\text{sym}}-9r^2\varepsilon \notag \\
\geq& r^2\left(1+\frac{1}{2}\sum_{\mu\neq\nu}V_{\mu}V_{\nu}+\frac{1}{3!}\sum_{\mu\neq\nu\neq\alpha\neq\mu}V_{\mu}V_{\nu}V_{\alpha}+(V_1V_2V_3V_4)_{\text{sym}}-9\varepsilon\right).
\end{align}
Using the relation \eqref{inequalityVmu}, we conclude that 
\begin{equation}
V_{\mu}V_{\nu}=(-V_{\mu})(-V_{\nu})\geq 0,
\end{equation}
so that each contribution to the two-product term as well as the four-product term is positive semidefinite. We use these properties and $1+V_{\mu} \geq 0$ to obtain
\begin{align}
- a^2 \nabla^2 +r^2 C^2 &> r^2\left(1+\frac{1}{2}\sum_{\mu\neq\nu}V_{\mu}V_{\nu}+\frac{1}{3!}\sum_{\mu\neq\nu\neq\alpha\neq\mu}V_{\mu}V_{\nu}V_{\alpha}-9
                          \varepsilon\right) \notag \\
  &> r^2\left(1+\frac{1}{3!}\sum_{\mu\neq\nu\neq\alpha\neq\mu}V_{\mu}V_{\nu}(V_{\alpha}+1)-9
    \varepsilon\right) \notag \\
&\geq r^2 \left(1- 9\varepsilon\right).
\end{align}

Using eq.\,\eqref{eq:M2bound}, we finally obtain
\begin{align}
- a^2 \nabla^2 + r^2  \left( \One (1 - s)-M^{(2)}  \right)^2
& = - a^2 \nabla^2 + r^2 C^2 - 2 r^2 (1 - s) M^{(2)} + r^2 (1 - s)^2
\mathbbm{1} \notag \\
&\geq r^2 \left(1- 9\varepsilon - 2| 1 - s | +| 1 - s |^2 \right)\notag \\
& = r^2 (1 - | 1 - s |)^2 - 9 r^2 \varepsilon
\end{align}
for $0<r\leq 1$. For the case $r>1$ we can decompose
\begin{equation}
\scalemath{0.9}{ -a^2 \nabla^2 + r^2  \left( \One(1 -s)- M^{(\mathsf {2})}  \right)^2
	=-a^2 \nabla^2 + \left( \One(1 - s)-M^{(2)}  \right)^2 + (r^2-1)  \left( \One(1 - s)-M^{(2)}\right)^2}\label{rgt1}
\end{equation}

and, since $r^2-1>0$, observe that the last term is positive
semidefinite. The first two terms, however, just correspond to the $r=1$
case, so the $r=1$ lower bound also applies for the $r>1$ case. All
together we thus have
\begin{equation}
  - a^2 \nabla^2 + r^2  \left( \One(1 - s)-M^{(\mathsf {2})}
  \right)^2
  >
  \left\{
    \begin{array}{ll}
      r^2 (1 - | 1 - s |)^2 - 9 r^2 \varepsilon  &  0<r\leq 1,\\
      (1 - | 1 - s |)^2 - 9  \varepsilon  &  r>1.
    \end{array}
  \right.
  \label{term132}
\end{equation}

\myparagraph{The second term}
As a result of eq.~\eqref{inequalitynabla} we find 
\begin{equation}
\left\Vert \sum_{\mu>\nu}\eta_{\mu}\eta_{\nu}a^{2}\left[\nabla_{\mu},\nabla_{\nu}\right]\right\Vert \leq\sum_{\mu>\nu}\left\Vert a^{2}\left[\nabla_{\mu},\nabla_{\nu}\right]\right\Vert < 6\varepsilon,
\end{equation}
so that we obtain the lower bound
\begin{equation}
-\sum_{\mu>\nu}\eta_{\mu}\eta_{\nu}a^{2}\left[\nabla_{\mu},\nabla_{\nu}\right]> -6\varepsilon\label{secondterm}
\end{equation}
for the second term.

\myparagraph{The fourth term}
From the commutation properties
\begin{equation}
C_{\mu}\eta_{\nu}=\begin{cases}
\eta_{\nu}C_{\mu} & \mu\geq\nu,\\
-\eta_{\nu}C_{\mu} & \mu<\nu,
\end{cases}
\end{equation}
it follows that
$C\eta_{\mu}=\left(-1\right)^{\mu+1}\eta_{\mu}C$. Similarly one can
show that
$\nabla_{\mu}\eta_{5}=\left(-1\right)^\mu\eta_{5}\nabla_{\mu}$. Using
these relations we find
\begin{align}
\left[M^{(2)},\eta_{\mu}\nabla_{\mu}\right]  &=\left(\epsilon\eta_{5}C\eta_{\mu}\nabla_{\mu}-\eta_{\mu}\nabla_{\mu} \epsilon\eta_{5}C\right)\notag \\
& =\epsilon\left(\eta_{5}C\eta_{\mu}\nabla_{\mu}+\eta_{\mu}\nabla_{\mu}\eta_{5}C\right)\notag \\
& =\epsilon\left(\eta_{5}\eta_{\mu}\left(-1\right)^{\mu+1}C\nabla_{\mu}+\eta_{5}\eta_{\mu}\left(-1\right)^\mu\nabla_{\mu}C\right)\notag \\
& =\epsilon \eta_{5}\eta_{\mu}\left(-1\right)^{\mu+1}\left[C,\nabla_{\mu}\right].\label{fourthterm}
\end{align}
From eqs.\,\eqref{clt1} and \eqref{inequalitynabla} we can then conclude that
\begin{align}
  \left\Vert a\left[M^{(2)},\eta_{\mu}\nabla_{\mu}\right]\right\Vert
  & \leq a\sum_{\mu}\left\Vert \left[C,\nabla_{\mu}\right]\right\Vert \notag \\
& \leq a\sum_{\mu\nu}\left\Vert \left[C_{\nu},\nabla_{\mu}\right]\right\Vert \notag \\
&< \sum_{\mu\neq\nu}\varepsilon \notag\\ 
&= 12\varepsilon
\end{align}
and thus we obtain the lower bound
\begin{equation}
  ar\left[M^{(2)},\eta_{\mu}\nabla_{\mu}\right]> -12r\varepsilon
  \label{term42}
\end{equation}
for all $r>0$.

\myparagraph{Final lower bound}
Combining eqs.\,\ref{term132}),\eqref{secondterm} and \eqref{term42}, we get a lower bound for
the two flavor operator %$A_2^{\dagger}A_2$
\begin{equation}
 A_2^{\dagger}A_2
  >
  \left\{
    \begin{array}{ll}
      r^2 (1 - | 1 - s |)^2 - (6+12r+9 r^2) \varepsilon  &  0<r\leq 1,\\
      (1 - | 1 - s |)^2 - (15+12r)  \varepsilon  &  r>1.
    \end{array}
  \right.\label{fibo2}
\end{equation}

\subsubsection{Lower bound on the one flavor operator $A_1^{\dagger}A_1$}

We will now try to find a lower bound on the operator 
\begin{equation}
A_{1}^{\dagger}A_{1}=-a^{2}\nabla^{2}-\sum_{\mu>\nu}\eta_{\mu}\eta_{\nu}a^{2}\left[\nabla_{\mu},\nabla_{\nu}\right]+r^{2}\left(2\cdot\One+M^{(1)}-s\One\right)^{2}+ar\left[M^{(1)},\eta_{\mu}\nabla_{\mu}\right], 
\end{equation}
by finding a bound of each term separately. Since the second term is
the same as in the two flavor case, we can take the previous result eq.\,\eqref{secondterm}. Once again, we consider the case $0<r\leq 1$ first.
\myparagraph{The first and third terms}

We start by observing that
\begin{align}
C_{\mu \nu}^2 & =  \frac{1}{4} (C_{\mu} C_{\nu} + C_{\nu} C_{\mu})^2 \notag\\
     & =  \frac{1}{4} (C_{\mu} C_{\nu} C_{\mu} C_{\nu} + C_{\mu} C_{\nu}
     C_{\nu} C_{\mu} + C_{\nu} C_{\mu} C_{\mu} C_{\nu} + C_{\nu} C_{\mu}
     C_{\nu} C_{\mu}) \notag\\
     & >  \frac{1}{4} (C_{\mu}^2 C_{\nu}^2 - \varepsilon + C_{\mu}^2
     C_{\nu}^2 - 2 \varepsilon + C_{\nu}^2 C_{\mu}^2 - 2 \varepsilon +
     C_{\nu}^2 C_{\mu}^2 - \varepsilon) \notag\\
     & =  \frac{C_{\mu}^2 C_{\nu}^2 + C_{\nu}^2 C_{\mu}^2 - 3
     \varepsilon}{2},
\end{align}
where we have used eq.~\eqref{inequalitynabla}. For $0 < r \leq
1$ we thus obtain the bound
\begin{align}
- a^2 \nabla^2 + r^2 (\mathbbm{1}+ M^{(1)})^2 & =  - a^2 \nabla^2 + r^2
     (1 + i \eta_{12} C_{12} + i \eta_{34} C_{34})^2 \notag\\
     & =  - \sum_{\mu} V_{\mu} + r^2 (C_{12}^2 + C_{34}^2 + \{ (1 + i
     \eta_{12} C_{12}), (1 + i \eta_{34} C_{34}) \} - 1) \notag\\
     & >  - \sum_{\mu} V_{\mu} + r^2 \left( \frac{C_1^2 C_2^2 +
     C_2^2 C_1^2 - 3 \varepsilon}{2} + \frac{C_3^2 C_4^2 + C_4^2 C_3^2 - 3
     \varepsilon}{2} - 1 \right) \notag\\
     & =  - \sum_{\mu} V_{\mu} + \frac{r^2}{2} (\{ (1 + V_1), (1 + V_2) \} +
     \{ (1 + V_3), (1 + V_4) \} - 2 - 6 \varepsilon) \notag\\
     & = - \sum_{\mu} V_{\mu} + \frac{r^2}{2} \left( 2 + 2 \sum_{\mu}
     V_{\mu} + \{ V_1, V_2 \} + \{ V_3, V_4 \} - 6 \varepsilon \right) \notag\\
     & \geq  - (1 - r^2) \sum_{\mu} V_{\mu} + r^2 (1 - 6 \varepsilon) \notag\\
     & \geq  r^2 - 6 r^2 \varepsilon.
\end{align}
For the general case of $0<s<2$ we use
\begin{equation}
\One+ M^{(1)} \geq - 1,
\end{equation}
which follows from $\left\| M^{(1)} \right\| \leq 2$, to find
\begin{align}
- a^2 \nabla^2 + r^2 \left( (2 - s) \mathbbm{1}+ M^{(1)}
\right)^2 & = - a^2 \nabla^2 + r^2 \left( (1 - s) \mathbbm{1}+ \left(
\mathbbm{1}+ M^{(1)} \right) \right)^2 \notag \\
& = - a^2 \nabla^2 + r^2 \left( \mathbbm{1}+ M^{(1)} \right)^2
+ r^2 (1 - s)^2 \mathbbm{1}+ 2 r^2 (1 - s) \left( \mathbbm{1}+ M^{(1)} \right) \notag\\
& >  r^2 - 6 r^2 \varepsilon + r^2 | 1 - s |^2 - 2 r^2 | 1 - s |\notag\\
& = r^2 (1 - | 1 - s |)^2 - 6 r^2 \varepsilon.
\end{align}
The lower bound of the first and third term for $0 <r\leq 1$ is thus given by
\begin{equation}
-a^2\nabla^2+r^2\left((2-s) \One +M^{(1)}\right)^2 > r^2 (1 - | 1 - s
|)^2 - 6 r^2 \varepsilon.
\end{equation}
For the $r>1$ case we can again show that the $r=1$ bound holds with
the same argument used in eq.\,\eqref{rgt1}. We thus obtain the general
lower bound
\begin{equation}
  -a^2\nabla^2+r^2\left((2-s) \One +M^{(1)}\right)^2
  >
  \left\{
    \begin{array}{ll}
      r^2 (1 - | 1 - s |)^2 - 6 r^2 \varepsilon  &  0<r\leq 1,\\
      (1 - | 1 - s |)^2 - 6  \varepsilon  &  r>1.
    \end{array}    \right.
  \label{term131}
\end{equation}

\myparagraph{The fourth term}
Let us first decompose the mass term
\begin{equation}
  a [M^{(1)}, \eta_{\mu} \nabla_{\mu}] = a \ii ( [\eta_{12} C_{12},
   \eta_{\mu} \nabla_{\mu}] +  [\eta_{34} C_{34}, \eta_{\mu} \nabla_{\mu}])
\end{equation}
and look at the first of the two commutators. We have
\begin{align}
a \ii [\eta_{12} C_{12}, \eta_{\mu} \nabla_{\mu}] & = a \ii (\eta_{12}
     [C_{12}, \eta_{\mu}] \nabla_2 + \eta_{\mu} \eta_{12} [C_{12},
     \nabla_{\mu}] + \eta_{\mu} [\eta_{12}, \nabla_{\mu}] C_{12})\notag\\
     & = a \ii (- 2 \eta_{12} \eta_2 C_{12} \nabla_2 + \eta_{\mu} \eta_{12}
     [C_{12}, \nabla_{\mu}] + 2 \eta_2 \eta_{12} \nabla_2 C_{12} )\notag\\
     & = a \ii (- 1)^{\delta_{\mu, 2}} \eta_{\mu} \eta_{12} [C_{12},
     \nabla_{\mu}],
\end{align}
which results in
\begin{align}
\| a \ii [\eta_{12} C_{12}, \eta_{\mu} \nabla_{\mu}] \| & =  \| a \ii (-
     1)^{\delta_{\mu, 2}} \eta_{\mu} \eta_{12} [C_{12}, \nabla_{\mu}] \|\notag\\
     & \leq  \frac{a}{2} (\| [C_1 C_2, \nabla_{\mu}] \| + \| [C_2 C_1,
     \nabla_{\mu}] \|)\notag\\
     & \leq  \frac{a}{2} (\| C_1 [C_2, \nabla_{\mu}] \| + \| [C_1,
     \nabla_{\mu}] C_2 \| + \| C_2 [C_1, \nabla_{\mu}] \| + \| [C_2,
     \nabla_{\mu}] C_1 \|) .
\end{align}
With eqs.\,\eqref{inequalitynabla} and \eqref{cleq1} we thus obtain the upper bound 
\begin{equation}
  \| a \ii [\eta_{12} C_{12}, \eta_{\mu} \nabla_{\mu}] \| < 2
  \varepsilon
\end{equation}
for the first term.
Similarly, we obtain for the second term
\begin{equation}
  \| a \ii [\eta_{34} C_{34}, \eta_{\mu} \nabla_{\mu}] \| = \| a \ii (-
   1)^{\delta_{\mu, 4}} \eta_{\mu} \eta_{34} [C_{34}, \nabla_{\mu}] \| < 2
   \varepsilon
 \end{equation}
and thus conclude
\begin{equation}
  a r [M^{(1)}, \eta_{\mu} \nabla_{\mu}] > - 4 r \varepsilon .
  \label{term41}
\end{equation}
\myparagraph{Final lower bound}
Combining eqs.\,\eqref{term131}, \eqref{secondterm} and \eqref{term41}, we get a lower bound for
the single flavor operator %$A_1^{\dagger}A_1$
\begin{equation}
 A_1^{\dagger}A_1
  >
  \left\{
    \begin{array}{ll}
      r^2 (1 - | 1 - s |)^2 - (6+4r+6 r^2) \varepsilon  &  0<r\leq 1\\
      (1 - | 1 - s |)^2 - (12+4r)  \varepsilon  &  r>1
    \end{array}
  \right.\label{fibo1}
\end{equation}

\section{Conclusion}
In this note we have proven that, when the admissibility condition
$\left\Vert \One-P\right\Vert <\varepsilon$  is imposed on every
plaquette $P$, both one and two flavor staggered overlap operators are
local. In particular, we can perform a Legendre expansion of the
inverse square root of $A^\dag A$, which is convergent if the spectral
condition of eq.\,\eqref{eq:MasterInequality} is fulfilled. From eqs.\,\eqref{fibo2}
and \eqref{fibo1}, we find that this is the case when
\begin{align}
  \varepsilon&< \frac{r^2 (1 - | 1 - s |)^2} {6+12r+9 r^2}  &  \text{two
                                                          flavor, }0<r\leq 1,\\
 \varepsilon&< \frac{(1 - | 1 - s |)^2}{15+12r}   & \text{two
                                                          flavor, } r>1,\\
                                               \varepsilon&<\frac{r^2 (1 - | 1 - s |)^2}{6+4r+6 r^2}  &
                                                          \text{single
                                                          flavor, }0<r\leq 1,\\
     \varepsilon&<\frac{(1 - | 1 - s |)^2}{12+4r}  & \text{single
                                                          flavor, } r>1,
\end{align}
which is dependent on the projection point $s$ and the Wilson
parameter $r$. The staggered overlap operator is thus conceptually on
the same footing as the standard overlap operator with a Wilson kernel.

\newpage

\begin{appendices}

\section{Plaquette dependent commutators}\label{AppendixA}

\subsection{Representations of the plaquette}

Since it is essential for the proof to have a bound on the plaquette,
we first want
to show how the plaquette can be represented. Let us define
the plaquette as the operator
\begin{equation}
  (P_{\mu \nu})_{xy} = U_{\mu} (x) U_{\nu} (x + \hat{\mu}) U_{\mu}^{\dag} (x +
\hat{\nu}) U_{\nu}^{\dag} (x) \delta_{x,y}.
\end{equation}
We find that
\begin{align}
(T_{\mu +} T_{\nu +} T_{\mu -} T_{\nu -})_{x y} & = U_{\mu} (x)
\delta_{x + \hat{\mu}, z} U_{\nu} (z) \delta_{z + \hat{\nu}, t}
U^{\dag}_{\mu} (u) \delta_{t - \hat{\mu}, u} U^{\dag}_{\nu} (y) \delta_{u
	- \hat{\nu}, y} \notag\\
& = U_{\mu} (x) U_{\nu} (x + \hat{\mu}) U^{\dag}_{\mu} (x + \hat{\nu})
U^{\dag}_{\nu} (y) \delta_{x, y} \notag\\
& = (P_{\mu \nu})_{x y}
\end{align}
or equivalently
\begin{equation}
P_{\mu \nu}  =T_{\mu +} T_{\nu +} T_{\mu -} T_{\nu -}.
\end{equation} 
Similarly, we can define plaquettes into negative coordinate directions as
\begin{align}
  P_{(- \mu) \nu} & = T_{\mu -} T_{\nu +} T_{\mu +} T_{\nu -}, \\
  P_{\mu (- \nu)} & = T_{\mu +}T_{\nu -} T_{\mu -} T_{\nu +}, \\
  P_{(- \mu )(- \nu)}& = T_{\mu -} T_{\nu -} T_{\mu +} T_{\nu +}.
\end{align}
With these, we can find commutation relations among the $T_{\mu\pm}$ ($\mu\neq\nu$) as
\begin{align}
[T_{\mu +}, T_{\nu +}] & = T_{\mu +} T_{\nu +} - T_{\nu +} T_{\mu
	+} \notag\\
& = T_{\mu +} T_{\nu +} (1 - T_{\nu -} T_{\mu -} T_{\nu +} T_{\mu
	+}) \notag\\
& = T_{\mu +} T_{\nu +} (1 - P_{(- \nu) (- \mu)})
\end{align}
and similarly for other combinations.

\subsection{Implications for some commutators}

We will need the commutator 
\begin{align}
a^2 [\nabla_{\mu}, \nabla_{\nu}]  & =\frac{1}{4}  ([T_{\mu +}, T_{\nu +}] + [T_{\mu -}, T_{\nu -}] -
[T_{\mu +}, T_{\nu -}] - [T_{\mu -}, T_{\nu +}])  \notag\\
& = \frac{1}{4} ( T_{\mu +} T_{\nu +} (1 - P_{(- \nu) (- \mu)}) + T_{\mu
	-} T_{\nu -} (1 - P_{\nu \mu}) \notag\\
&   - T_{\mu +} T_{\nu -} (1 - P_{\nu (- \mu)}) - T_{\mu -}
T_{\nu +} (1 - P_{(- \nu) \mu}) ),
\end{align}
where we used eq.\,(\ref{Nabla-mu}). Imposing a smoothness condition
\begin{equation}
\| \One - (P_{\mu\nu})_{xx} \| < \varepsilon
\end{equation}
on every plaquette
and remembering that all $\| T_{\mu \pm} \| = 1$, we find that
\begin{align}
a^2 \| [\nabla_{\mu}, \nabla_{\nu}] \| &<
\frac{\varepsilon}{4} (\| T_{\mu +} T_{\nu +} \| + \| T_{\mu -} T_{\nu -}
\| + \| T_{\mu +} T_{\nu -} \| + \| T_{\mu -} T_{\nu +} \|) \notag\\
& = \varepsilon.
\end{align}
Similarly we find
\begin{align}
[C_{\mu}, C_{\nu}] & = \frac{1}{4} ([T_{\mu +}, T_{\nu +}] + [T_{\mu
	-}, T_{\nu +}] + [T_{\mu +}, T_{\nu -}] + [T_{\mu -}, T_{\nu -}])\\
& = \frac{1}{4} (T_{\mu +} T_{\nu +} (1 - P_{(- \nu) (- \mu)}) + T_{\mu -}
T_{\nu +} (1 - P_{(- \nu) \mu}) \notag\\
&  + T_{\mu +} T_{\nu -} (1 - P_{\nu (- \mu)}) + T_{\mu -}
T_{\nu -} (1 - P_{\nu \mu}))
\end{align}
and thus
\begin{equation}
\| [C_{\mu}, C_{\nu}] \| < \varepsilon.
\end{equation}
Using the fact that $\| C_{\mu} \| \leq 1$, we can also infer that
\begin{equation}
\left\| [C_{\mu}, C_{\nu}] \prod_{i = 1}^n C_{\alpha_i} \right\| <
\varepsilon
\end{equation}
for any number $n$ of additional $C_{\alpha}$ terms. We thus see that
\begin{equation}
\| C^2 - (C_1^2 C_2^2 C_3^2 C_4^2) _{\text{sym}}\| < N \varepsilon,
\end{equation}
 where $N$ is determined by the number of commutations we have to perform to
bring the terms in $C^2$ into the correct order. Let us first rewrite
\begin{equation}
  C^2 - (C_1^2 C_2^2 C_3^2 C_4^2)_{\text{sym}}=
 \frac{1}{4!}P_{\alpha\beta\gamma\delta} (C_\alpha C_\beta C_\gamma C_\delta
 C-C_\alpha ^2 C_\beta ^2 C_\gamma ^2 C_\delta ^2) .
\end{equation}
For each term in the symmetrization bracket we now perform the
commutations in two
steps. First we bring the terms in $C$ into order, so we are left with
$(C_\alpha C_\beta C_\gamma C_\delta)^2$. For each of the $4!$ products
in $C$ this requires a different number of commutations, namely
\begin{align*}
\text{Number of commutations}  &: \quad 0 \quad 1 \quad 2 \quad 3 \quad 4 \quad 5 \quad 6\\
\text{Number of products} &: \quad 1 \quad 3 \quad 5 \quad 6 \quad 5 \quad 3 \quad 1
\end{align*}
On average we thus have 3 commutations in this first step. From there
on it takes 6 more commutations to obtain $C_\alpha ^2 C_\beta ^2
C_\gamma ^2 C_\delta^2$, so we have performed 9 commutations on
average. Since we average over all permutations, we have
\begin{equation}
\| C^2 - (C_1^2 C_2^2 C_3^2 C_4^2) _{\text{sym}}\| < 9 \varepsilon.
\end{equation}
In order to find $a\|[C_{\mu},\nabla_{\nu}]\|$ we use eqs.\,(\ref{Cmu}) and (\ref{Nabla-mu}) to determine
\begin{align}
a[C_{\mu},\nabla_{\nu}]&=\frac{1}{4}([T_{\mu+}, T_{\nu+}] + [T_{\mu-}, T_{\nu+}] - [T_{\mu+}, T_{\nu-}] - [T_{\mu-}, T_{\nu-}]) \notag\\
& = \frac{1}{4} (T_{\mu+} T_{\nu+} (1 - P_{(- \nu) (- \mu)}) + T_{\mu-}
T_{\nu+} (1 - P_{(- \nu) \mu}) \notag\\
&  - T_{\mu+} T_{\nu-} (1 - P_{\nu (- \mu)}) - T_{\mu-}
T_{\nu-} (1 - P_{\nu \mu})),
\end{align}
from which it follows that
\begin{equation}
\|a[C_{\mu},\nabla_{\nu}]\| < \varepsilon.
\end{equation}
Also, for $\mu=\nu$ the commutator trivially vanishes.
\end{appendices}

\bibliographystyle{apsrev4-2}

\bibliography{literature}

\end{document}